\documentclass[a4paper,fleqn,usenatbib,useAMS]{mnras}
\usepackage{graphicx}   
\usepackage{color}
\usepackage{lscape}
\usepackage{txfonts}
\usepackage{amsfonts}

\title[Thermal soft X-ray component in NSs]
{The advection-dominated accretion flow for the origin of the thermal soft X-ray component in  
low-level accreting neutron stars}
\author[Erlin Qiao and B.F. Liu]{Erlin Qiao $^{1,2}$\thanks{E-mail:
qiaoel@nao.cas.cn} and B.F. Liu $^{1,2}$\\
$^{1}$Key Laboratory of Space Astronomy and Technology, National Astronomical Observatories, Chinese Academy of
Sciences, Beijing 100012, China \\
$^{2}$School of Astronomy and
Space Sciences, University of Chinese Academy of Sciences, 19A Yuquan Road, Beijing 100049, China\\} 

\date{Accepted XXX. Received YYY; in original form ZZZ}
\pubyear{2018}

\begin{document}
\label{firstpage}
\pagerange{\pageref{firstpage}--\pageref{lastpage}}
\maketitle
\begin{abstract}
A thermal soft X-ray component is often detected in low-level accreting neutron stars (NSs), 
but is not detected in low-level accreting stellar-mass black holes (BHs). 
In this paper, we investigate the origin of such a thermal soft X-ray component in the 
framework of the self-similar solution of the advection-dominated accretion flow (ADAF) around NSs.
It is assumed that a fraction, $f_{\rm th}$, of the energy transferred onto the surface of the NS
is thermalized at the surface of the NS as the soft photons to be scattered in the ADAF.
We self-consistently calculate the structure and the corresponding emergent spectrum
of the ADAF by considering the radiative coupling between the soft photons from the surface 
of the NS and the ADAF itself. We show that the Compton $y$-parameter of the ADAF for NSs
is systematically lower than that of BHs. Meanwhile, we find that the temperature of the thermal 
soft X-ray component in NSs decreases with decreasing mass accretion rate, which is qualitatively 
consistent with observations. We test the effect of $f_{\rm th}$ on the structure, as well as the 
emergent spectrum of the ADAF. It is found that a change of $f_{\rm th}$ can significantly change 
the temperature of the thermal soft X-ray component as well as the spectral slope in hard X-rays. 
Finally, it is suggested that the value of $f_{\rm th}$ can be constrained by fitting the high-quality 
X-ray data, such as the $\it XMM$-$\it Newton$ spectrum between 0.5-10 keV in the future work.
\end{abstract}




\section{Introduction}
Low-mass X-ray binaries (XRBs) are accreting systems with either a stellar-mass black hole (BH)  
or a neutron star (NS) drawing gas from its low-mass companion star 
($\lesssim 1 M_{\odot}$) via its Roche lobe. Most of the low-mass XRBs are  X-ray transients, which
are discovered when they first go into outburst. The outburst can last from a few weeks to a few months, then
decayed into quiescence for months to years \citep[][]{Tanaka1995,Remillard2006}.
Generally, there are two dominated spectral states in both BHXRBs and NSXRBs, 
i.e., the high/soft spectral  state with a relatively higher mass accretion rate, 
and the low/hard spectral state with a relatively lower mass accretion rate
\citep[][]{Done2007,Gilfanov2010}. 
The soft-to-hard state transition occurs at about $1-4\%$ of the  
Eddington luminosity \citep[][]{Maccarone2003}, while for some sources, the hysteresis is observed, i.e.,
the transition luminosity of the soft-to-hard state transition is $\sim $ 3 times lower than of the 
transition luminosity of the hard-to-soft transition 
\citep[e.g.][]{Maccarone2003b,Nowak2002,Kubota2004,Rodriguez2003,
MeyerHofmeister2005,Gladstone2007,Zhangzl2016,Zhanghui2018}.
So far, a great of effects have been made for exploring the accretion physics in the
different spectral states of XRBs. Currently, it is widely believed that in the high/soft state, 
the accretion is dominated by the cool, geometrically thin, optically thick disc \citep[][]{Shakura1973}.
While in the low/hard spectral state, the accretion is dominated by the hot,  
optically thin, advection-dominated accretion flow (ADAF)
\citep[][for review]{Narayan1994,Narayan1995,Yuan2014}.

Theoretically, for the ADAF around astrophysical BHs, a fraction of the viscously dissipated energy 
in the ADAF will be advected into the event horizon of the BHs rather than being radiated out
\citep[][]{Ichimaru1977,Rees1982,Narayan1994,Abramowicz1995,Chen1995}. 
The fraction of the energy advected into the black hole depends on the mass accretion rate.  
Specifically, for a relatively higher mass accretion rate within the ADAF regime, such as 
the mass accretion rate close to the rate corresponding to the spectral state transition in BHXRBs, 
the fraction of the energy advected into the black hole is small, and 
the radiative efficiency of the ADAF can be nearly comparable with that of the cool disc \citep[][]{Xie2012}. 
The radiative efficiency of the ADAF decreases rapidly with decreasing mass accretion rate $\dot M$. 
Especially, if $\dot M \lesssim 5\times 10^{-4} \dot M_{\rm Edd}$ 
(with $\dot M_{\rm Edd}$ = $1.39 \times 10^{18} M/M_{\rm \odot} \rm \ g \ s^{-1}$), 
the accretion is completely dominated by the advection, and the accretion flow is radiatively
inefficient \citep[][]{Mahadevan1997}. 
While for the ADAF around  NSs, due to the existence of a hard surface of NSs, it is expected that 
the energy  advected onto the surface of the NSs will eventually be radiated out, so the accretion flow 
is always radiately efficient. 
The observational data indeed support that the quiescent BH accreting systems are fainter than that of 
the quiescent NS accreting systems with similar mass accretion rates, as can be seen in the 
diagram between the luminosity and the binary orbital period 
\citep[e.g.][]{Menou1999,Lasota2000,Garcia2001,Hameury2003,McClintock2004}. 
The study of \citet[][]{Narayan1995} showed that there exists a critical mass 
accretion rate $\dot M_{\rm crit}$. For $\dot M \gtrsim \dot M_{\rm crit}$, the accretion flow will 
transit from the hot ADAF  to the cool disc. 
For NS accreting systems, the energy advected onto the surface of the NS will eventually radiate out,
which in turn can effectively cool the accretion flow, resulting in the accretion 
flow  much easier to collapse. So theoretically the critical mass accretion rate  
$\dot M_{\rm crit}$ for NSs is generally lower than that of BHs. 
Specifically, the critical mass accretion rate is $\dot M_{\rm crit}\sim \alpha^2 \dot M_{\rm Edd}$
(with $\alpha$ being the viscosity parameter)
for a  BH accreting system, while the critical mass accretion rate is 
$\dot M_{\rm crit}\sim 0.1\alpha^2 \dot M_{\rm Edd}$ for a NS accreting system \citep[][]{Narayan1995}. 

Observationally, in low-level accreting NSs with the $0.5-10$ keV X-ray luminosity  
in the range of about $10^{34}$ to $10^{36}$  erg $\rm s^{-1}$ corresponding to $0.01\%-1\%$ of 
the Eddington luminosity, a thermal soft X-ray component 
is detected \citep[e.g.][]{Jonker2004,ArmasPadilla2013a,ArmasPadilla2013b,Degenaar2013,Campana2014}.
While such a thermal soft X-ray component is not detected in accreting BHs in the 
similar luminosity range \citep[e.g.][]{Wijnands2015}.
In some low-level accreting NSs , it is found that the temperature of such a thermal 
soft X-ray component decreases with decreasing the X-ray luminosity, which implies that 
the origin of such a thermal soft X-ray component in low-level accreting NSs is accretion-related 
\citep[][]{ArmasPadilla2013c,Degenaar2013,Bahramian2014}. 
By fitting the X-ray spectrum of a sample composed of twelve BH and NS XRBs, it is found that  
the electron temperature  $T_{\rm e}$ of the corona around NSs 
is $\sim 15-25$ keV, while the electron temperature $T_{\rm e}$ is $\sim 30-200$ keV around BHs. 
Meanwhile, it is found that the Compton $y$-parameter for NSs is systemically lower than that 
of BHs \citep[][]{Burke2017}. The lower electron temperature and the lower Compton $y$-parameter 
of the corona for NSs compared with the case for BHs are believed to be resulted by the strong Compton 
cooling of the corona by the soft thermal photons from the surface of the NSs
\citep{Sunyaev1989pre,syunyaev1991,Narayan2008}.

The very faint X-ray binary IGR J17062-6143 discovered in 2006 is believed to be a NSXRB due to the
detection of a Type-I X-ray burst in 2012 \citep[][]{Degenaar2013burst}.
By fitting the spectral energy distribution (SED) of IGR J17062-6143, it was shown that the thermal soft X-ray 
component can not originate from the accretion disc \citep[][]{HernandezSantisteban2018}. 
\citet[][]{vandenEijnden2018} fitted  the soft X-ray data of IGR J17062-6143 in three different epochs, 
i.e., 2014 ($\it Chandra$), 2015 ($\it Swift$), and 2016 ($\it XMM$-$\it Newton$), suggesting that
the thermal soft X-ray component is from the surface of the NS with a radius $\sim 11-12$ km. 
Furthermore, it is found that the temperature of the thermal soft X-ray component decreases 
from $0.48\pm 0.01 \rm keV$ (2014) and $0.47\pm 0.01 \rm keV$ (2015) in the first two epochs 
to  $0.36\pm 0.01 \rm keV$ (2016) with $\sim 2$ times decrease of the X-ray luminosity 
\citep[][]{vandenEijnden2018}, which are qualitatively consistent with the theoretical 
predictions of the low-level accretion around NSs \citep[][]{Zampieri1995}.

In this paper, we investigate the origin of such a thermal soft X-ray component in low-level accreting 
NSs based on the self-similar solution of the ADAF \citep[][]{Narayan1995}. Specifically,   
we consider that the internal energy stored in the ADAF and the radial kinetic 
energy of the ADAF are transfered onto the surface of the NS. A fraction, $f_{\rm th}$, of this energy 
is assumed to be thermalized as the blackbody emission, which is then scattered in the ADAF. 
We self-consistently calculate the structure of the ADAF by considering the coupling of the blackbody 
emission from the surface of the NS and the ADAF itself. With the derived structure of the ADAF, we adopt
the multi-scattering method to calculate the emergent of the ADAF around NSs. 
We compare the structure and the emergent spectrum of the ADAF between BHs and NSs.  We study the 
structural and the spectral features of the ADAF with the mass accretion rates, especially, the 
temperature of the thermal soft X-ray component with the mass accretion rates. We test the effect 
of $f_{\rm th}$ on the structural and the spectral features of the ADAF with a focus on the  
relationship between the temperature of the thermal soft X-ray component and $f_{\rm th}$.
The model is briefly introduced in Section 2. The numerical results are shown in Section 3. 
The discussions are in Section 4, and the conclusions are in Section 5.

\section{The model}
In this paper, we calculate the structure of the ADAF around a NS or a BH based on 
the self-similar solution of the ADAF \citep[][]{Narayan1995}. For clarity, we list the 
equations as follows.

Equation of state,
\begin{eqnarray}\label{state}
p_{\rm g}=\beta \rho c_{\rm s^2}={ {\rho k T_{\rm i}}\over {\mu_{\rm i} m_{\rm p}} }
+{ {\rho k T_{\rm e}} \over {\mu_{\rm e}m_{\rm p}} }, 
\end{eqnarray}
where $p_{\rm g}$ is the gas pressure, $m_{\rm p}$ is the proton mass, and $\beta$ is the magnetic parameter  
(with magnetic pressure $p_{\rm m}=B^2/{8\pi}=(1-\beta)p$, $p
=p_{\rm g}+p_{\rm m}$), $T_{\rm i}$ is the ion temperature and $T_{\rm e}$
is the electron temperature, $\rho$ is the density, $k$ is the Boltzmann constant,
$\mu_{\rm i}$ and $\mu_{\rm e}$ are the effective molecular weights of ions and 
electrons respectively, which can be expressed as,
\begin{eqnarray}\label{weight}
\mu_{\rm i}={4\over {1+3X}}=1.23,\ \ \  \mu_{\rm e} ={2\over {1+X}}=1.14,
\end{eqnarray}
where the hydrogen mass fraction $X=0.75$ is adopted for the numerical values 
of $\mu_{\rm i}$ and $\mu_{\rm e}$. The internal energy per unit volume of the gas is,
\begin{eqnarray}\label{internal}
U={3\over 2}p_{\rm g} + {B^2\over {4\pi}}.
\end{eqnarray}

In the following, we list the radial velocity $\rm v$, the angular velocity $\Omega$, the isothermal 
sound speed $c_{\rm s}$, the density $\rho$, the magnetic field $B$, the pressure $p$,
the electron number density $n_{\rm e}$, the viscous dissipation of energy per unit volume $q^{+}$,
and the scattering optical depth $\tau_{\rm es}$ in the vertical direction derived from 
the self-similar solution of the ADAF, which are all functions of $m$, $\dot m$, $\alpha$,
$\beta$ and $r$ \citep[][]{Narayan1995},    
\begin{eqnarray}\label{self}
\begin{array}{l}
{\rm v}= -2.12\times10^{10}\alpha c_{1}r^{-1/2} \ \ \ \rm cm \ s^{-1},    \\
\Omega  = 7.19\times10^{4}c_{2}m^{-1}r^{-3/2} \ \ \ \rm s^{-1}, \\
c_s^2 =4.50\times10^{20}c_{3}r^{-1} \ \ \rm cm^{2} \ s^{-2}, \\
\rho  =3.79\times10^{-5}\alpha^{-1}c_{1}^{-1}c_{3}^{-1/2}m^{-1}\dot m r^{-3/2} \ \ \ \rm g\ cm^{-3}, \\
p  =1.71\times10^{16}\alpha^{-1}c_{1}^{-1}c_{3}^{1/2}m^{-1}\dot m r^{-5/2} \ \ \ \rm g\ cm^{-1}\ s^{-2}, \\
B = 6.55\times10^8\alpha^{-1/2}(1-\beta)^{1/2}c_{1}^{-1/2}c_{3}^{1/4}
  m^{-1/2}\dot m^{1/2}r^{-5/4} \ \ \ \rm G,    \\
n_e=\rho/{\mu_{\rm e}m_{\rm p}}=2.00\times10^{19}\alpha^{-1}c_1^{-1}c_{3}^{-1/2}m^{-1} 
    \dot m_{\rm } r^{-3/2}\ \ \ \rm cm^{-3},  \\
q^{+}=1.84\times 10^{21}\varepsilon^{'}c_{3}^{1/2}m^{-2}\dot m_{\rm } r^{-4}\ \ \rm ergs \ cm^{-3} \ s^{-1}, \\
\tau_{\rm es}  = 2n_{e}\sigma_{T}H = 12.4\alpha^{-1}c_{1}^{-1}\dot m r^{-1/2},
\end{array}
\end{eqnarray}
where $m$ is the  mass of a NS or a BH scaled with the solar mass $M_{\rm \odot}$, $\dot m$ is 
the accretion rate scaled with the Eddington accretion rate $\dot M_{\rm Edd}$, $r$ is the radius 
from the black hole scaled with the Schwarzschild radius 
$R_{\rm S}$ (with $R_{\rm S}=2.95\times 10^5 m \ \rm {cm}$), and  
\begin{equation}\label{coef}
\begin{array}{l}
{c_1}={(5+2\varepsilon^{'}) \over {3\alpha^2}}g(\alpha,\varepsilon^{'}),\\
\\
{c_3}={2\varepsilon(5+2\varepsilon^{'})\over {9\alpha^2} } g(\alpha,\varepsilon^{'}),\\
\\
{\varepsilon{'}}={\varepsilon\over f}={1\over f} \biggl({{5/3-\gamma}\over {\gamma-1}}\biggr),\\
\\
g(\alpha,\varepsilon^{'})=\biggl[ {1+{18\alpha^2\over (5+2\varepsilon^{'})^{2}}\biggr]^{1/2}-1},\\
\\
\gamma={{32-24\beta-3\beta^2}\over {24-21\beta}},
\end{array}
\end{equation}
with $f$ being the advected fraction of the viscously dissipated energy. 
The energy balance of the ADAF is determined by the following equations, 
\begin{equation}\label{energy}
\begin{array}{l}
q^{+}=fq^{+}+q^{\rm ie}\\
q^{\rm ie}=q^{-}\\
\end{array}
\end{equation}
where $q^{\rm ie}$ is the energy transfer rate from ions to electrons via  Coulomb collision
\citep[]{Stepney1983}, which is given by, 
\begin{equation}\label{qie}
\begin{array}{l}
q_{\rm ie}=3.59\times 10^{-32}n_{\rm e}n_{\rm i}(T_{\rm i}-T_{\rm e}){ {{1+T^{'1/2}}} \over T^{'{3/2}} }, 
\end{array}
\end{equation}
with 
\begin{equation}\label{qie}
\begin{array}{l}
\end{array}
T^{'}={kT_{\rm e}\over {m_{\rm e}c^2}}\biggl(1+{m_{\rm e}\over m_{\rm p}} {T_{\rm i}\over T_{\rm e}}\biggr), 
\end{equation}
and $m_{\rm e}$ is the electron mass, $c$ is the speed of light.
$q^{-}=q^{-}_{\rm brem}+ 
q^{-}_{\rm syn}+ 
q^{-}_{\rm brem, C}+ 
q^{-}_{\rm syn, C}+
q^{-}_{\rm *, C}$ 
is the cooling rate of the 
electrons with $q^{-}_{\rm brem}$, $q^{-}_{\rm syn}$, $q^{-}_{\rm brem, C}$, $q^{-}_{\rm syn, C}$
and $q^{-}_{\rm *,C}$ being the bremsstrahlung 
cooling rate, the synchrotron cooling rate, the Compton cooling rate by  self-Comptonization
of bremsstrahlung radiation,  the Compton cooling rate by  self-Comptonization
of synchrotron radiation, and the Compton cooling rate by Comptonization
of the radiation from the central NS. 
The expressions for $q^{-}_{\rm brem}$, $q^{-}_{\rm syn}$, $q^{-}_{\rm brem, C}$, $q^{-}_{\rm syn, C}$
are all the standard expressions as in \citep[][]{Narayan1995}. 

Here for NSs, we consider that the dominated soft photons from the surface of the NS to be scattered 
in the accretion flow are from the accretion itself. Specifically, we consider that the internal energy and 
the radial kinetic energy of the ADAF was transfered onto the surface of the NS. A fraction $f_{\rm th}$ 
of this energy is assumed to be thermalized as the blackbody emission, which is then scattered in the  
ADAF. The energy of the ADAF transfered onto the 
surface of the NS per second  can be expressed as, 
\begin{equation}\label{soft_L}
\begin{array}{l}
L_{*} = 4\pi R_{*}H(R_{*})|{\rm v}(R_{*})|\left[U(R_{*})+{1\over2}\rho(R_{*}){\rm v}^2(R_{*})\right],
\end{array}
\end{equation}
where $R_{*}$ is the radius of the NS, and $U(R_{*})$ is the internal energy of the gas 
at $R_{*}$ as equation (\ref{internal}), $H(R_{*})$ is the scaleheight of the gas at $R_{*}$,  
$\rho(R_{*})$ and  ${\rm v}(R_{*})$ are the density of the gas and the radial velocity of the gas 
at $R_{*}$ respectively as in equation (\ref{self}). 
In this case, if the radiation from the surface of the NS is assumed to be isotropic, 
the effective temperature of the radiation $T_{*}$ can be given by,
\begin{equation}\label{internal-e}
\begin{array}{l}
T_{*}=\bigl({L_{*}f_{\rm th}\over {4\pi R_{*}^{2}\sigma}}\bigr)^{1/4}, 
\end{array}
\end{equation}
where $\sigma$ is the Stefan-Boltzmann constant. The outgoing flux from the NS reaching at a 
radius R is given by, 
\begin{equation}\label{soft_F}
\begin{array}{l}
F_{*}(R)=\bigr({L_{*}\over {4\pi R^2}}\bigr) {e}^{-\tau^{'}_{\rm es}(R)},
\end{array}
\end{equation}
where $\tau^{'}_{\rm es}(R)$ is the scattering optical depth in the radial direction from the surface 
of the NS. Such a flux $F_{*}(R)$ at a distance $R$ is locally scattered in the ADAF, and the
Compton cooling rate $q^{-}_{*,C}$ is then derived accordingly. 
One can refer to equation (3.28-3.30) in \citep[]{Narayan1995} for details.
$q^{-}_{*, C}$ is assumed to be zero for BHs due to the existence of the 
event horizon. 

Substituting the formulae of $\rho$ and $c^2_{\rm s}$ in equation (\ref{self})
into equation (\ref{state}), the equation of state of the gas can be re-expressed as, 
\begin{equation}\label{state1}
\begin{array}{l}
T_{\rm i}+1.08T_{e}=6.66\times10^{12}\beta c_{3}r^{-1}.
\end{array}
\end{equation}
We solve equations (\ref{energy}) and (\ref{state1}) for the ion temperature $T_{\rm i}$,
electron temperature $T_{\rm e}$ and the advected fraction of the viscously dissipated energy
$f$ by specifying the black  hole mass $m$, accretion rate $\dot m$, viscosity parameter $\alpha$, 
magnetic parameter $\beta$ and $f_{\rm th}$ describing the fraction of the internal energy 
and the radial kinetic energy transfered onto the surface of the NS
to be thermalized as the blackbody emission.
With the derived electron temperature $T_{\rm e}$ and the scattering optical depth $\tau_{\rm es }$  in the 
vertical direction, we calculate the corresponding emergent spectrum of the ADAF around a NS
with the multi-scattering of  photons in the hot gas. One can refer to \citet[][]{Qiao2010,Qiao2013} or
\citet[][]{Manmoto1997} for the calculation of the emergent spectrum for details.

\section{Numerical results}
We calculate the structure of the ADAF by specifying $m$, $\alpha$, $\beta$ and $f_{\rm th}$.
Throughout the paper, we take $m=10$ and $m=1.4$ for BH and NS respectively. Meanwhile, we take 
the inner boundary of the ADAF as $3R_{\rm S}$ for BHs (non-rotating) and 12.5 km 
\footnote{Assuming the central star with the mass $M$ and radius $R_{*}$, the
gravitational potential energy released by the accretion of a unit mass to its surface is
$\Delta E=GM/R_{*}$. For a non-rotating black hole, $R_{*}$ is assumed to be as $3R_{\rm S}$,
so $\Delta E=c^2/6$.  In this paper, in order to keep the same energy release by the 
accretion of a unit mass between the BH and NS, if $m=1.4M_{\odot}$ is adopted for a NS,
the corresponding radius of a NS is $R_{*}=$ 12.5 km.} 
for NSs. we set $\alpha=0.3$ as usual. The magnetic field in the ADAF solution is relatively weak, as 
suggested by magnetohydrodynamic simulations \citep[][]{Yuan2014}.
We fix $\beta=0.95$ in this paper. 

\subsection{Black hole vs. Neutron star}\label{sec_BH}
\begin{figure*}
\includegraphics[width=85mm,height=60mm,angle=0.0]{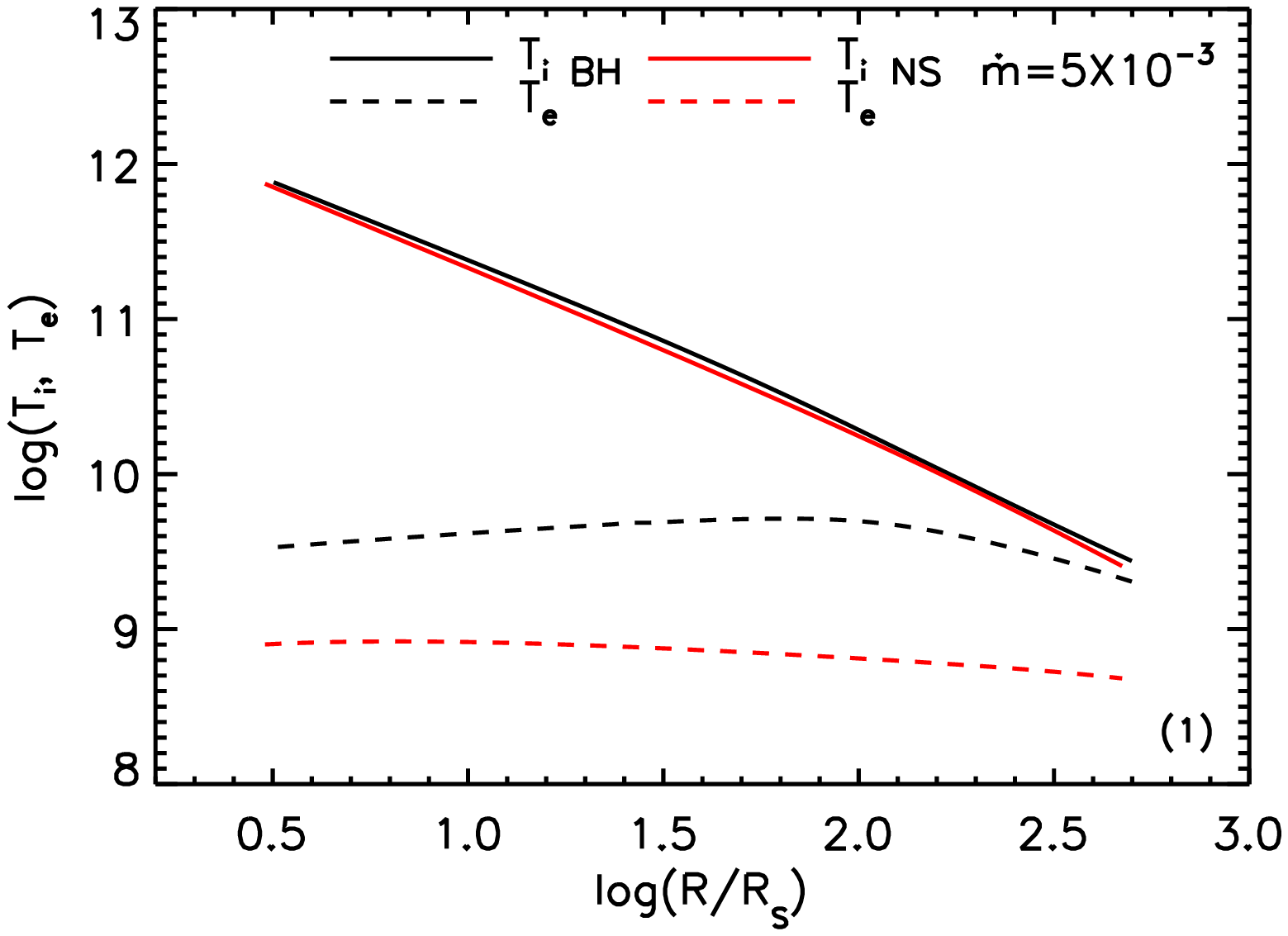}
\includegraphics[width=85mm,height=60mm,angle=0.0]{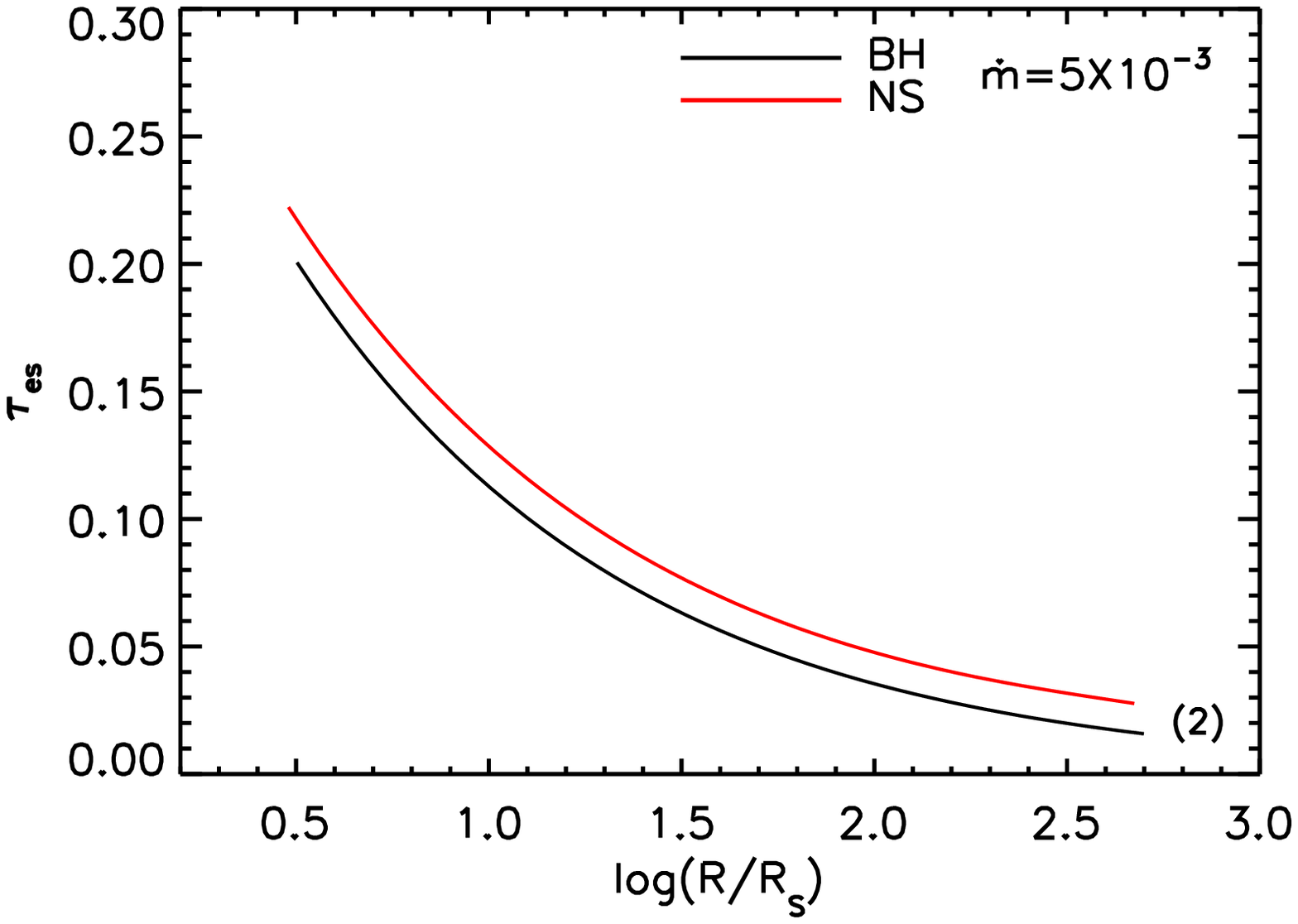}
\includegraphics[width=85mm,height=60mm,angle=0.0]{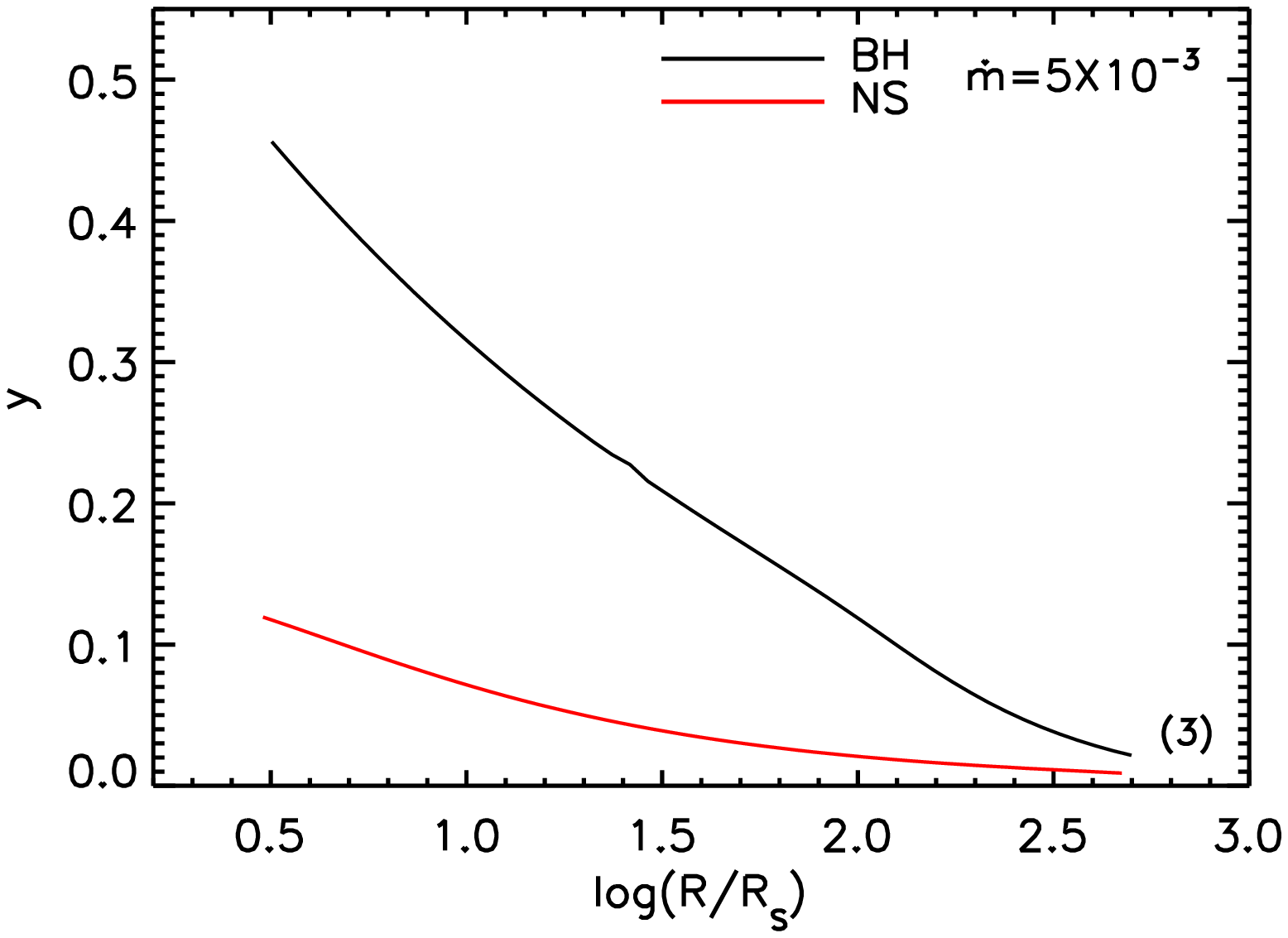}
\includegraphics[width=85mm,height=60mm,angle=0.0]{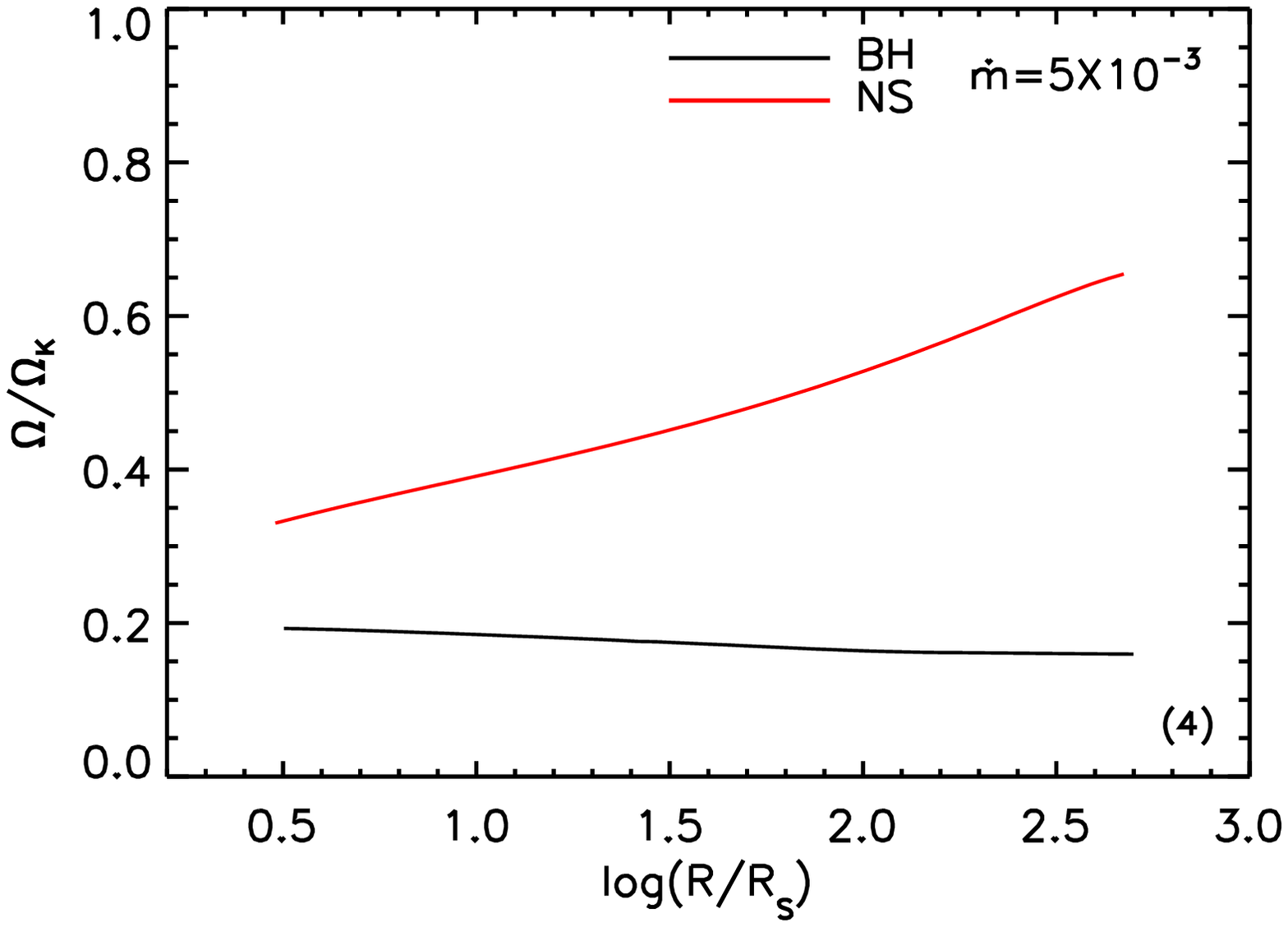}
\caption{\label{bh-ns}Panel (1): ion temperature $T_{\rm i}$ (solid line) as a function of radius and 
electron temperature $T_{\rm e}$ (dashed line) as a function of radius. The black lines are for 
BHs and the red lines are for NSs.  Panel (2): Compton scattering optical depth $\tau_{\rm es}$ 
as a function of radius. The black line is for BHs and the red line is for NSs.
Panel (3): Compton $y$-parameter as a function of radius.  
The black line is for BHs and the red line is for NSs.
Panel (4): $\Omega/\Omega_{\rm K}$ as a function of radius. 
The black line is for BHs and the red line is for NSs.}
\end{figure*}

\begin{figure*}
\includegraphics[width=85mm,height=60mm,angle=0.0]{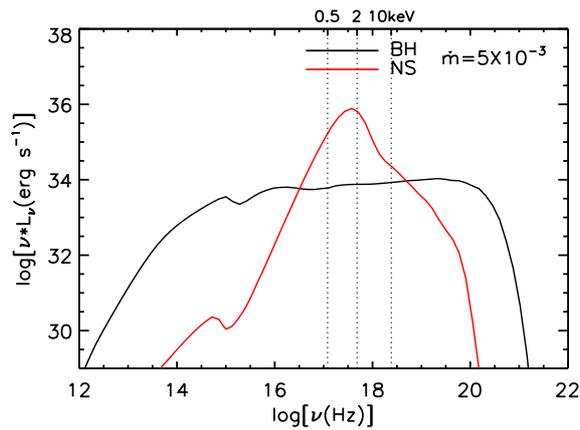}
\caption{\label{sp:bh-ns} Emergent spectra of the ADAF.  
The black line is for BHs and the red line is for NSs. }
\end{figure*}

In the panel (1) of Fig. (\ref{bh-ns}), we plot the ion temperature $T_{\rm i}$
and the electron temperature $T_{\rm e}$ of the ADAF as a function of radius 
for BHs and NSs respectively with $\dot m=5\times 10^{-3}$. $f_{\rm th}=1$ is adopted for NSs.
The black-solid line and the black-dashed line are the ion temperature and the electron temperature 
of the ADAF for BHs respectively.  The red-solid line and the red-dashed line are the ion temperature 
and the electron temperature of the ADAF for NSs respectively.
It is clear that the ion temperature of BHs and NSs is very similar, which intrinsically can be predicted 
by the basic assumptions set in the ADAF, i.e., the ions are first heated by the viscous process, and
then a fraction of this heat energy stored in the ions is transfered to the elections via
Coulomb collision. Since the density of the gas in the ADAF is low,  
the Coulomb collision is not efficient, the fraction of the heat energy transfered to the electrons is small.   
Consequently, the ions can keep a relatively higher temperature, which is close to the virial temperature.
We note that the electron temperature of BHs is $\sim$ 3 times higher than that of NSs,
This can be understood as the strong Compton cooling of the electrons in the ADAF by the soft photons 
from the surface of the NSs, which however does not exist for BHs. 
In the panel (2) of  Fig. (\ref{bh-ns}), we plot the Compton scattering optical depth 
$\tau_{\rm es}$ as a function of radius for BHs and NSs respectively with $\dot m=5\times 10^{-3}$. 
$f_{\rm th}=1$ is adopted for NSs. The black line is for BHs and the red line is for NSs. 
It is shown that the Compton scattering optical depth for BHs is slightly less than that of NSs.
In the panel (3) of  Fig. (\ref{bh-ns}), we plot the Compton $y$-parameter
(defined as $y={{4kT_{\rm e}}\over {m_{\rm e}c^2}} \rm {Max}(\tau_{\rm es}, \tau^2_{\rm es})$) 
as a function of radius for BHs and NSs respectively with $\dot m=5\times 10^{-3}$. 
$f_{\rm th}=1$ is adopted for NSs. The black line is for BHs and the red line is for NSs. 
It is found that the Compton $y$-parameter for BHs is significantly larger than that of 
NSs, which intrinsically can predict a harder hard X-ray spectrum for BHs compared with NSs.
In the panel (4) of  Fig. (\ref{bh-ns}), we plot the ratio of the angular velocity
of the ADAF to the Keplerian angular velocity, $\Omega/\Omega_{\rm K}$ 
(with $\Omega_{\rm K}$ being the Keplerian angular velocity), 
as a function of radius for BHs and NSs respectively with $\dot m=5\times 10^{-3}$. 
$f_{\rm th}=1$ is adopted for NSs.  
The black line is for BHs and the red line is for NSs. 
It can be seen that $\Omega/\Omega_{\rm K}$ for BHs is systematically lower than that of NSs.  
As discussed in \citep[][]{Narayan1994} for the self-similar solution of the ADAF, 
$\Omega \approx {[2\epsilon^{'}/{(5+2\epsilon^{'})}]}^{1/2} \Omega_{\rm K}$
(with $ \epsilon^{'}\propto 1/f$). 
For the case of very efficient cooling, $f\rightarrow 0$, 
$\epsilon^{'}\rightarrow \infty$, and $\Omega \rightarrow \Omega_{\rm K}$.
Meanwhile, it is clear that $\Omega$ decreases with increasing $f$. 
The radiative efficiency of the ADAF around BHs is intrinsically lower than that of
NSs, resulting in a lower value of  $\Omega/\Omega_{\rm K}$ for BHs compared with the case for NSs. 
It is also found that $\Omega/\Omega_{\rm K}$ increases with increasing radius for NSs, indicating  
that the angular velocity of the ADAF is closer to the  
Keplerian value in the outer region. 
One can refer to Fig. (\ref{sp:bh-ns}) for the corresponding emergent spectra, 
as the black line is for BHs and the red line is for NSs.
In Table (\ref{mass_effect}), we list the related radiative features of the ADAF 
around BHs and NSs respectively. As an example, for a NS with  
$\dot m=5\times 10^{-3}$ and $f_{\rm th}=1$, the temperature of the thermal soft X-ray component
from the surface of the NS is $0.45$ keV, and the X-ray luminosity between 0.5 to 10 keV is 
$1.0\times 10^{36}$ erg $\rm s^{-1}$. While for a BH with 
$\dot m=5\times 10^{-3}$, due to the existence of the event horizon, all the 
internal energy stored in the ADAF and radial kinetic energy of the ADAF at $R_{*}$ are advected into the  
event horizon without radiation. So no such a thermal component is predicted for BHs, as 
can be seen from the emergent spectrum in Fig. (\ref{sp:bh-ns}). 
The X-ray luminosity between 0.5 to 10 keV for BHs is $2.2\times 10^{34}$ erg $\rm s^{-1}$
nearly two orders of magnitude lower than that of NSs, which is also one of the
basic conclusions predicted by the ADAF solution, i.e., most of the viscously dissipated 
energy is advected into the event horizon of the BH without radiation, while the energy advected onto the 
surface of the NS will be eventually radiated out. 

\begin{table*}
\caption{Radiative features of the ADAF around BHs and NSs. 
$T_{*}$ is the effective temperature at the surface of NSs.
$L_{\rm 0.5-10 keV}$ is the luminosity between 0.5 and 10 $\rm keV$.}
\centering
\begin{tabular}{ccccccccc}
\hline\hline
$m$ & $R_{*} $  & $\alpha$ & $\beta$ & $f_{\rm th}$ & $\dot m$  & $T_{*} \ (\rm keV)$ & 
$L_{\rm 0.5-10 keV}\ (\rm erg \ s^{-1}) $ \\
\hline
$10 \ ($\rm BH$)$ & 3$R_{\rm S}$ &0.3  &0.95   &1.0   &  $5\times10^{-3}$     &-        & $2.2\times 10^{34}$ \\
$1.4\ ($\rm NS$)$ & 12.5 (km)    &0.3  &0.95   &1.0   &  $5\times10^{-3}$     &0.45    & $1.0\times 10^{36}$ \\
\hline\hline
\end{tabular}
\\
\label{mass_effect}
\end{table*}

\subsection{ The dependence of the structure and emergent spectra on $\dot m$ }
\begin{figure*}
\includegraphics[width=85mm,height=60mm,angle=0.0]{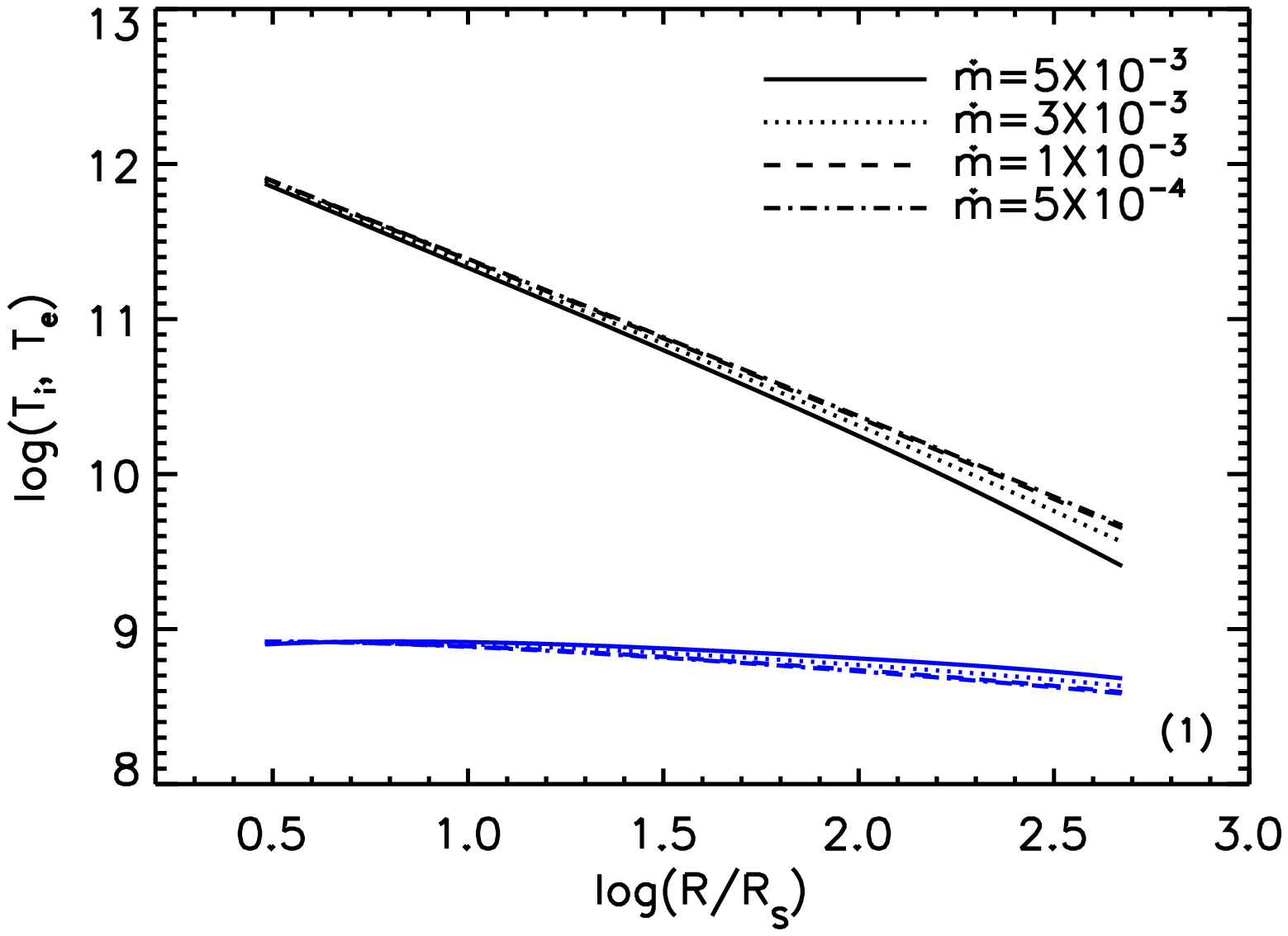}
\includegraphics[width=85mm,height=60mm,angle=0.0]{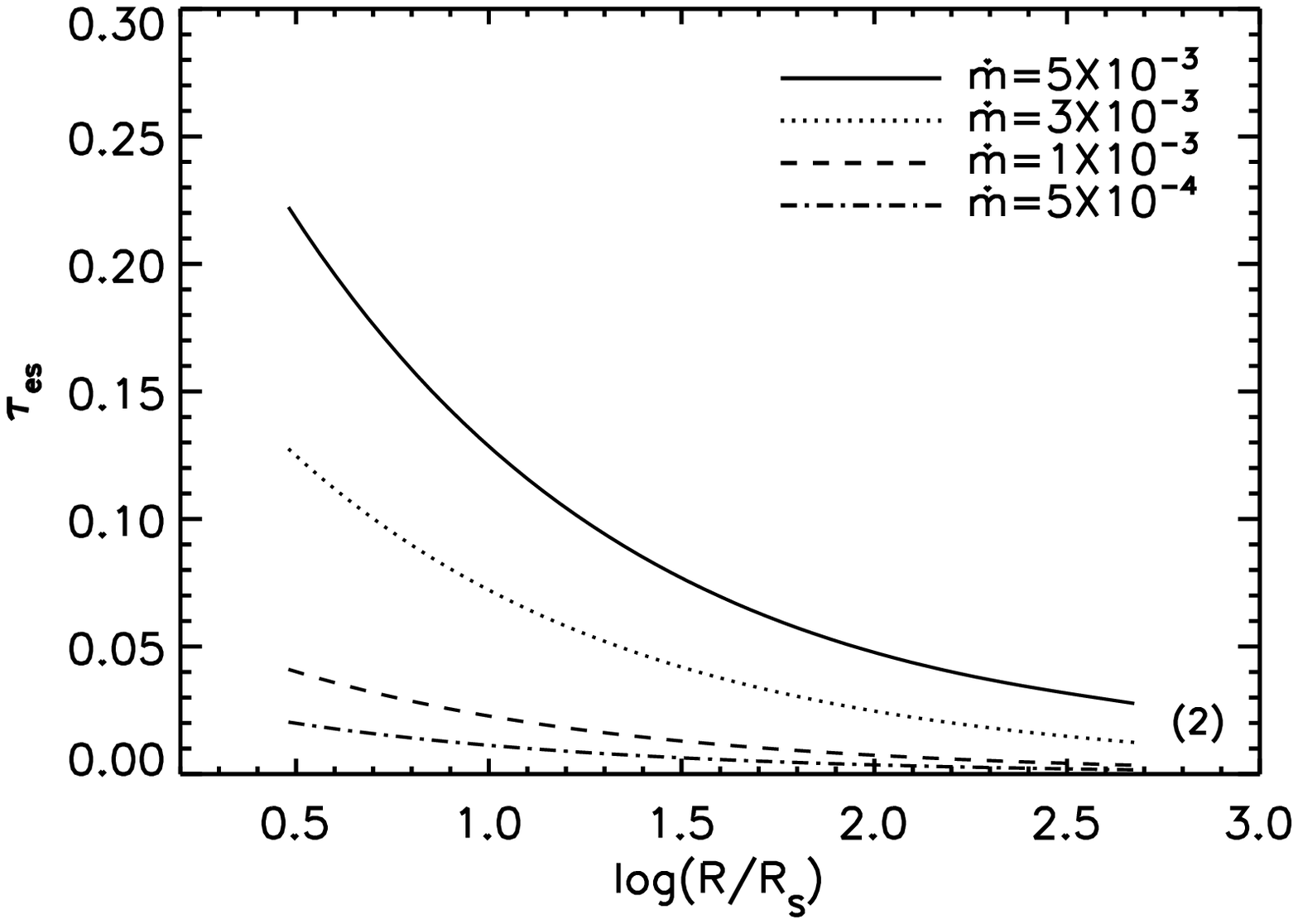}
\includegraphics[width=85mm,height=60mm,angle=0.0]{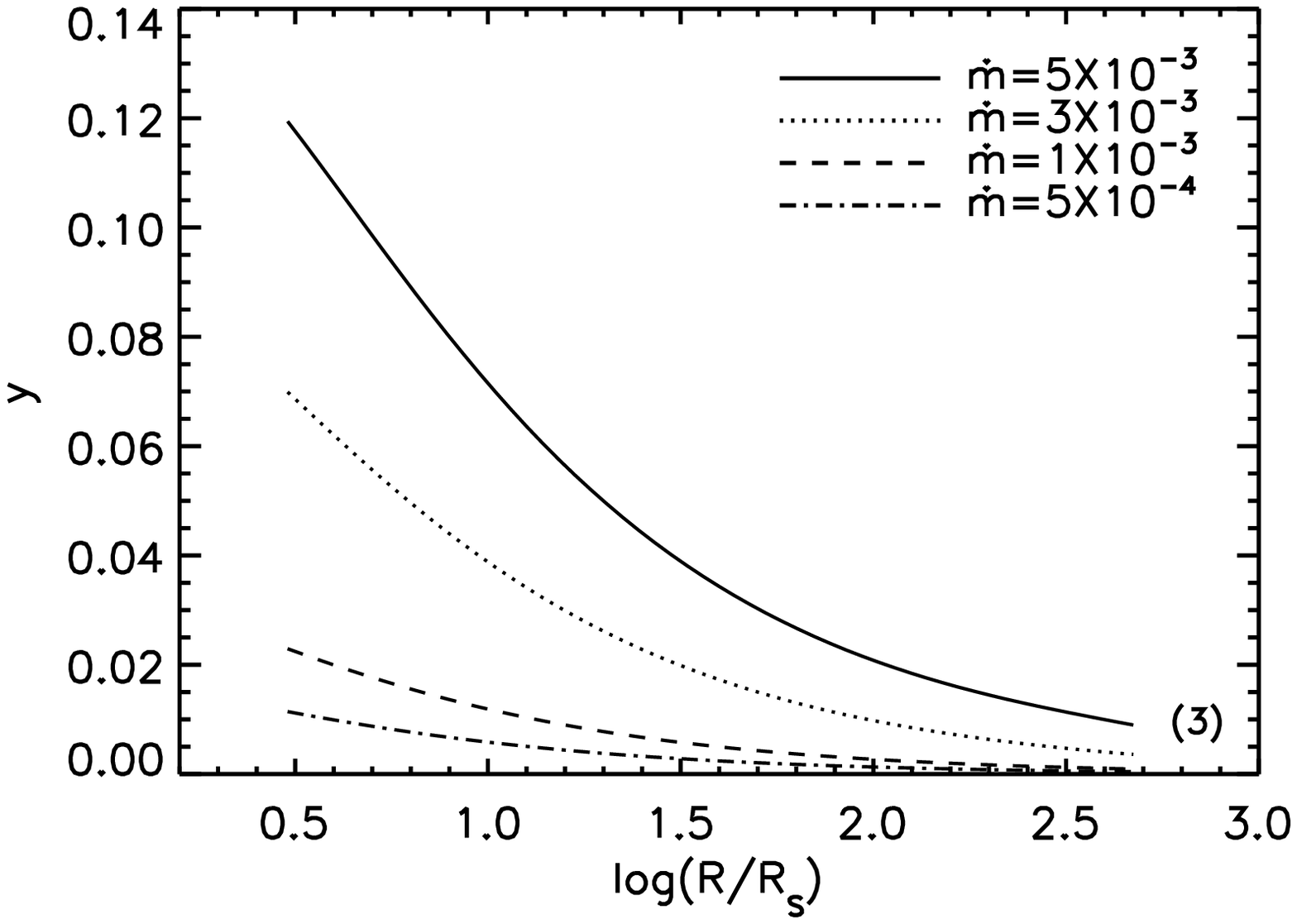}
\includegraphics[width=85mm,height=60mm,angle=0.0]{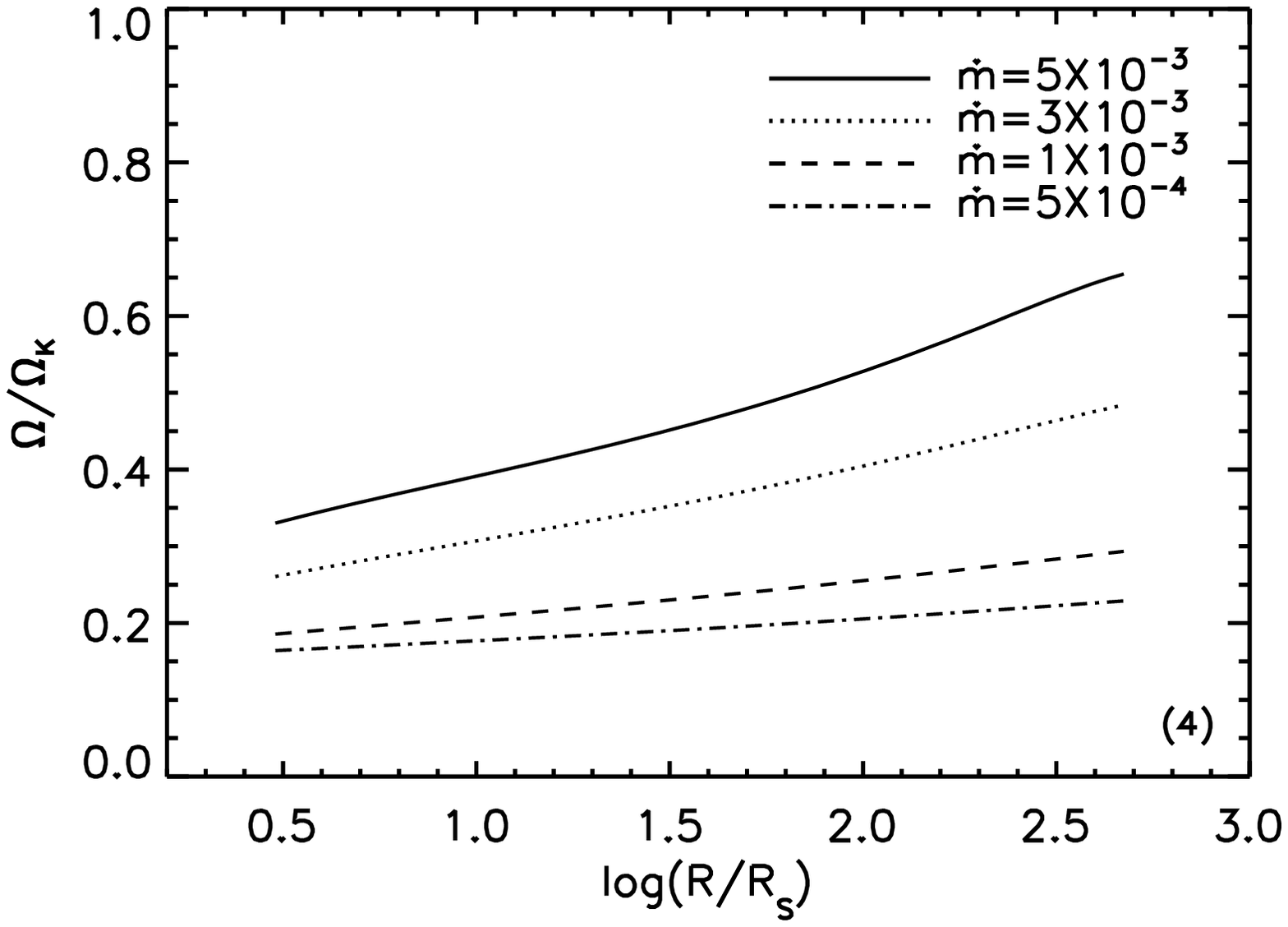}
\caption{\label{mdot}Panel (1): ion temperature $T_{\rm i}$ (black line) as a function of radius and 
electron temperature $T_{\rm e}$ (blue line) as a function of radius around NSs for different $\dot m$
with $f_{\rm th}=1$.
Panel (2): Compton scattering optical depth $\tau_{\rm es}$ as a function of radius around NSs
for different $\dot m$ with $f_{\rm th}=1$.
Panel (3): Compton $y$-parameter as a function of radius around NSs for different 
$\dot m$ with $f_{\rm th}=1$.
Panel (4): $\Omega/\Omega_{\rm K}$ as a function of radius around 
NSs for different $\dot m$ with $f_{\rm th}=1$.} 
\end{figure*}

\begin{figure*}
\includegraphics[width=85mm,height=60mm,angle=0.0]{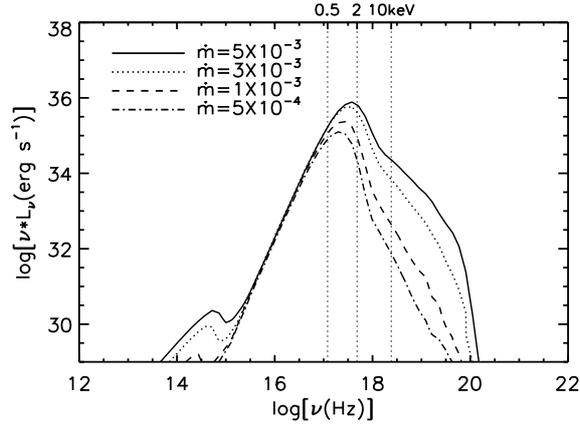}
\caption{\label{sp:mdot} Emergent spectra of the ADAF around NSs for different 
$\dot m$ with $f_{\rm th}=1$. } 
\end{figure*}

In the panel (1) of Fig. \ref{mdot}, we plot the ion temperature $T_{\rm i}$
and the electron temperature $T_{\rm e}$ of the ADAF as a function of radius  for 
NSs for different $\dot m$ with $f_{\rm th}=1$. It can be seen that both 
$T_{\rm i}$ and $T_{\rm e}$ change very slightly with $\dot m$ around NSs, which is very
similar to the case of BHs \citep[][]{Mahadevan1997}. 
In the panel (2) of  Fig. (\ref{mdot}), we plot the Compton scattering optical depth 
$\tau_{\rm es}$ as a function of radius for NSs for different $\dot m$ with $f_{\rm th}=1$. 
The Compton scattering optical depth $\tau_{\rm es}$ systemically decreases with 
decreasing $\dot m$, as predicted by the formula
of $\tau_{\rm es}$ in equation (\ref{self}).
In the  panel (3) of  Fig. (\ref{mdot}),  we plot the Compton $y$-parameter as a function of 
radius for NSs for different $\dot m$ with $f_{\rm th}=1$.
It is clear that the Compton $y$-parameter decreases with decreasing $\dot m$, which predicts 
a softer hard X-ray spectrum with decreasing $\dot m$, as can be seen in  Fig. (\ref{sp:mdot}). 
In the  panel (4) of  Fig. (\ref{mdot}),  we plot $\Omega/\Omega_{\rm K}$ 
as a function of radius for NSs for different $\dot m$ with $f_{\rm th}=1$. 
It can be seen that $\Omega/\Omega_{\rm K}$ systematically  decreases with decreasing $\dot m$, 
which can be roughly understood as, with a decrease of $\dot m$, the radiative efficiency 
of the ADAF slightly decreases even in the case of NSs, resulting in 
a decrease of the angular velocity, as has been discussed in Section \ref{sec_BH}. 
In Table (\ref{mdot_effect}), we list the radiative features of the ADAF around NSs 
for different $\dot m$ with $f_{\rm th}=1$. 
It can be seen that the ratio of the energy of the ADAF transfered onto the surface of the 
NS per second, $L_{*}$, to the accretion luminosity, $L_{\rm G}$, (defined as $L_{\rm G}=GM\dot M/R_{*}$)
slightly increases with decreasing $\dot m$, which however does not affect the trend that  
the temperature of the thermal soft X-ray 
component decreases from  $0.45$ keV to $0.26$ keV for the mass accretion rate decreasing from 
$\dot m=5\times 10^{-3}$ to $\dot m=5\times 10^{-4}$.
Meanwhile, the X-ray luminosity between 0.5 to 10 keV decreases from 
$1.0\times 10^{36}$ erg $\rm s^{-1}$ to $1.3\times 10^{35}$ erg $\rm s^{-1}$ 
for the mass accretion rate decreasing from 
$\dot m=5\times 10^{-3}$ to $\dot m=5\times 10^{-4}$.
It is interesting to note that, a linear correlation between log $L_{\rm 0.5-10 keV}$ and  
log $\dot m$ is found, i.e., 
\begin{equation}\label{liner}
\begin{array}{l}
{\rm log}\ L_{\rm 0.5-10keV}=38.1+0.9\times {\rm log}\ \dot m, 
\end{array}
\end{equation}
which can be simply expressed as $L_{\rm 0.5-10keV} \propto \dot m^{0.9}$.
The fitting result of $L_{\rm 0.5-10keV} \propto \dot m^{0.9}$ for the ADAF around NSs is very close to 
the predictions by the radiative efficient accretion flows as generally,  
$L \propto \dot m$. For the ADAF around BHs, theoretically, ADAF is radiatively inefficient and    
$L \propto \dot m^s$, with $ {\rm s} \sim 2$ for $5\times 10^{-3} \lesssim \dot m \lesssim 2\times 10^{-2}$,  
with $ {\rm s} \sim 1.6$ for $10^{-4} \lesssim \dot m \lesssim 5\times 10^{-3}$, and 
with $ {\rm s} \sim 3.4$ for $\dot m \lesssim 10^{-4}$ \citep[][]{Merloni2003}.

\begin{table*}
\caption{Radiative features of the ADAF around NSs for different $\dot m$. 
$L_{*}/L_{\rm G}$ is the ratio of the energy of the ADAF transfered onto the 
surface of the NS per second to the accretion luminosity. 
$T_{*}$ is the effective temperature at the surface of NSs.
$L_{\rm 0.5-10 keV}$ is the luminosity between 0.5 and 10 $\rm keV$.}
\centering
\begin{tabular}{ccccccccc}
\hline\hline
$m$ & $R_{*} \ ($\rm km$)$  & $\alpha$ & $\beta$ & $f_{\rm th}$ & $\dot m$  & $L_{*}/L_{\rm G}$  & $T_{*} \ (\rm keV)$ 
& $L_{\rm 0.5-10 keV}\ (\rm erg \ s^{-1}) $ \\
\hline
$1.4$ & 12.5 &0.3  &0.95   & 1.0  &  $5\times10^{-3}$     &56.0\%    &0.45  &  $1.0\times 10^{36}$ \\
$1.4$ & 12.5 &0.3  &0.95   & 1.0  &  $3\times10^{-3}$     &58.6\%    &0.40  &  $7.5\times 10^{35}$ \\
$1.4$ & 12.5 &0.3  &0.95   & 1.0  &  $1\times10^{-3}$     &60.8\%    &0.31  &  $2.7\times 10^{35}$ \\
$1.4$ & 12.5 &0.3  &0.95   & 1.0  &  $5\times10^{-4}$     &61.3\%    &0.26  &  $1.3\times 10^{35}$ \\
\hline\hline
\end{tabular}
\\
\label{mdot_effect}
\end{table*}

\subsection{The effect of $f_{\rm th}$}\label{sec_fth} 
\begin{figure*}
\includegraphics[width=85mm,height=60mm,angle=0.0]{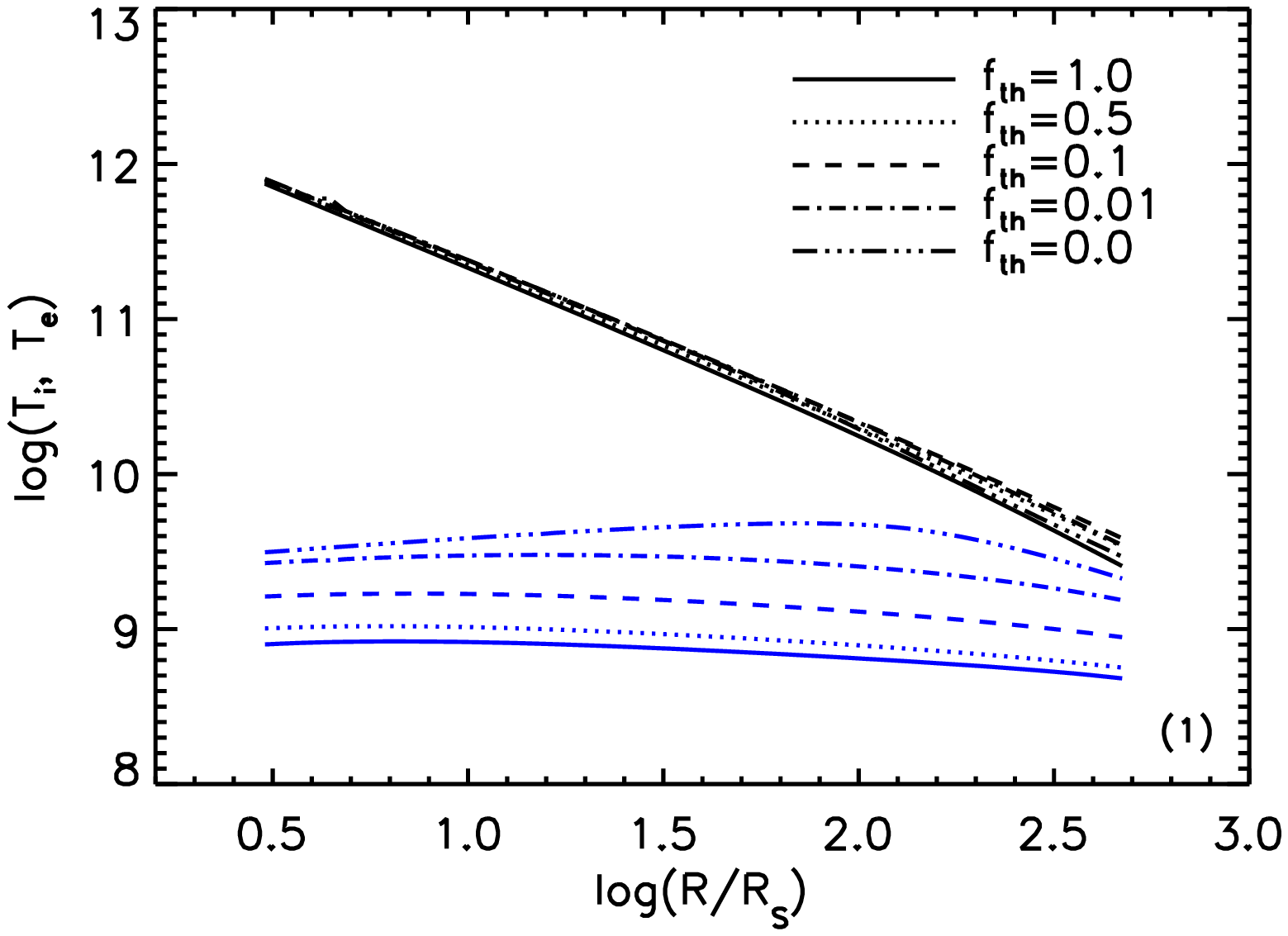}
\includegraphics[width=85mm,height=60mm,angle=0.0]{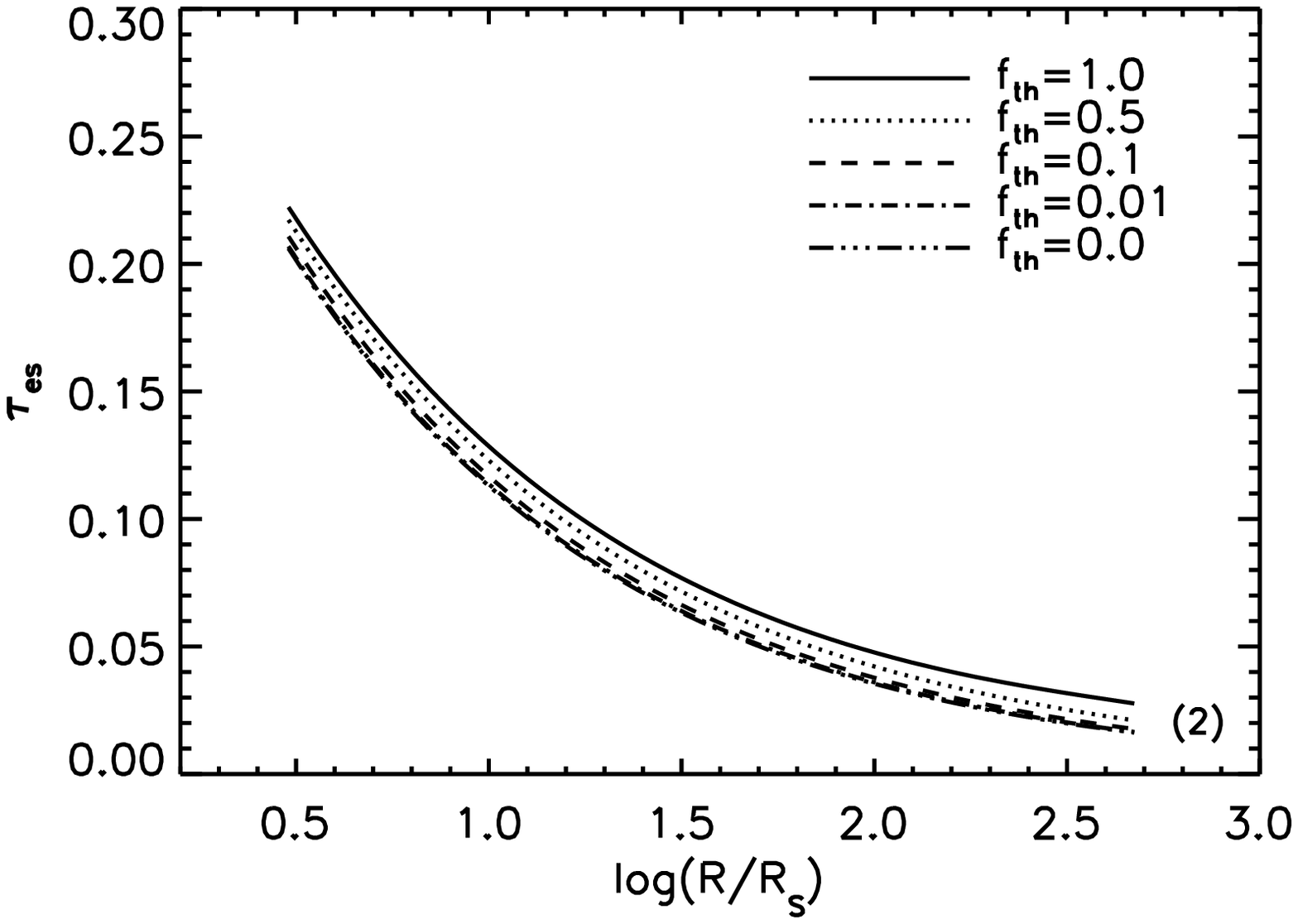}
\includegraphics[width=85mm,height=60mm,angle=0.0]{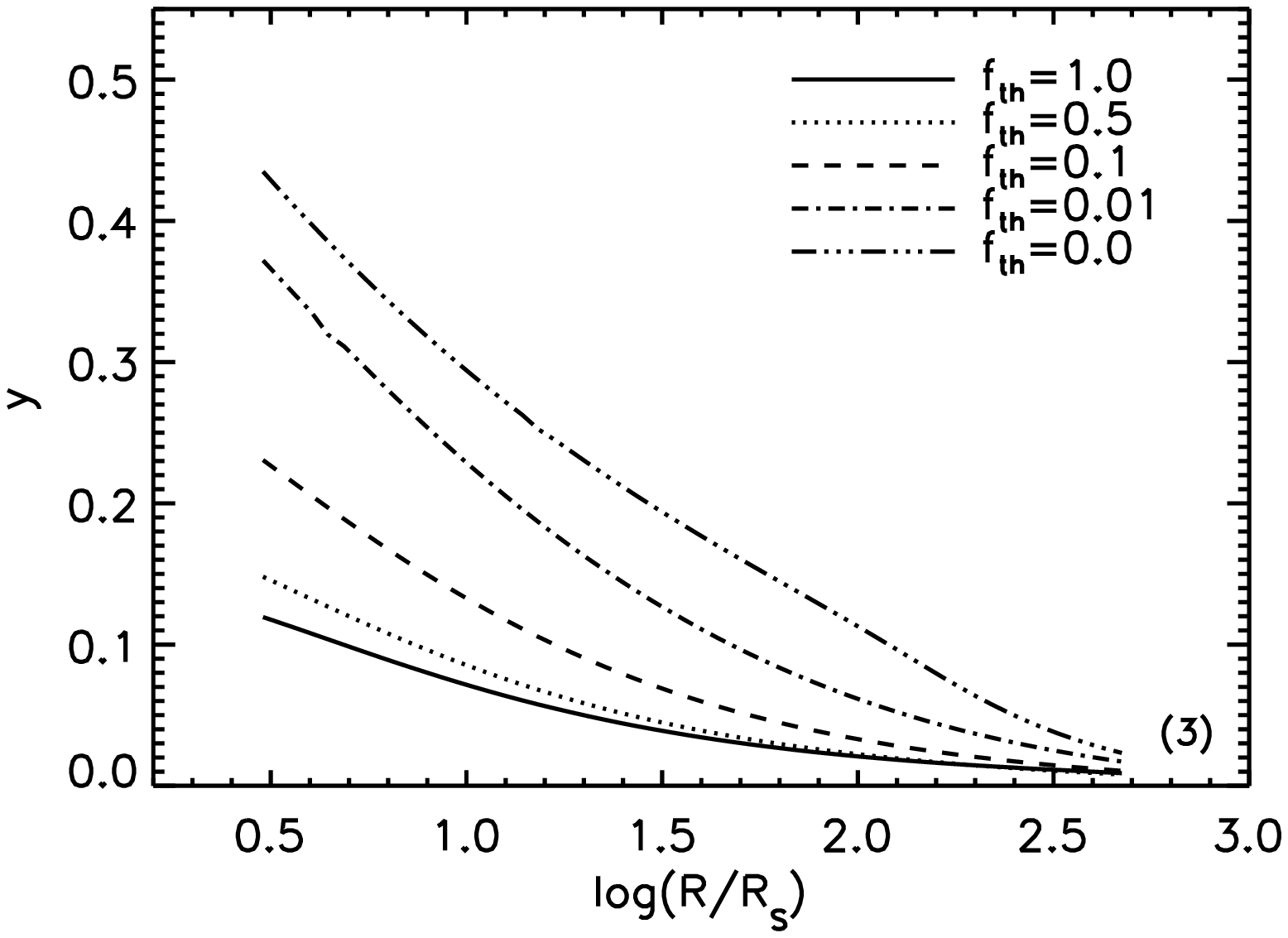}
\includegraphics[width=85mm,height=60mm,angle=0.0]{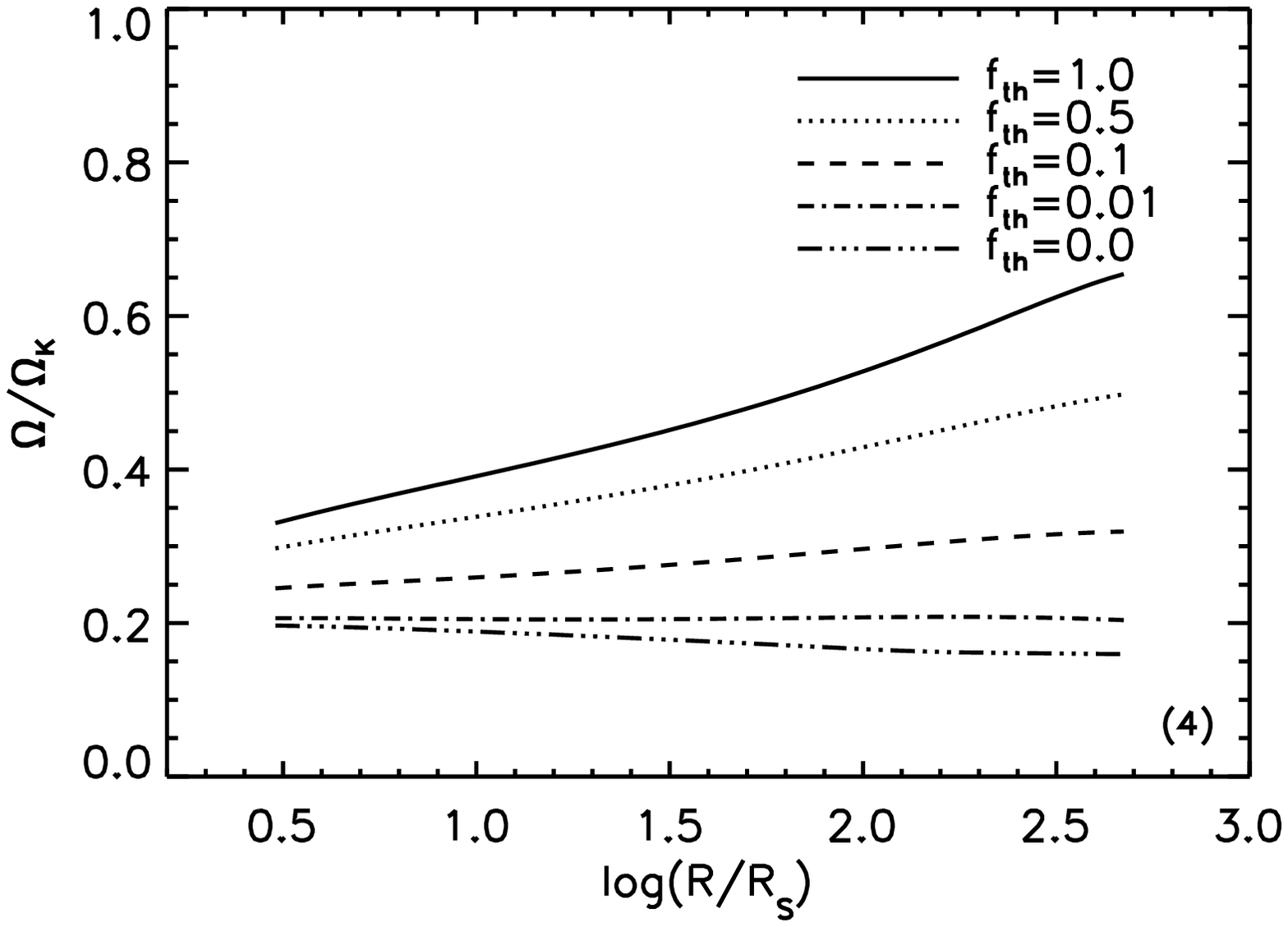}
\caption{\label{fth} Panel (1): ion temperature $T_{\rm i}$ (black line) and 
electron temperature $T_{\rm i}$ (blue line) as a function of radius around NSs for different $f_{\rm th}$ 
with $\dot m=5\times 10^{-3}$.
Panel (2): Compton scattering optical depth $\tau_{\rm es}$ as a function of radius around NSs
for different $f_{\rm th}$ with $\dot m=5\times 10^{-3}$.
Panel (3): Compton $y$-parameter as a function of radius around NSs for different 
$f_{\rm th}$ with $\dot m=5\times 10^{-3}$.
Panel (4): $\Omega/\Omega_{\rm K}$ as a function of radius around NSs 
for different $f_{\rm th}$ with $\dot m=5\times 10^{-3}$. }
\end{figure*}

\begin{figure*}
\includegraphics[width=85mm,height=60mm,angle=0.0]{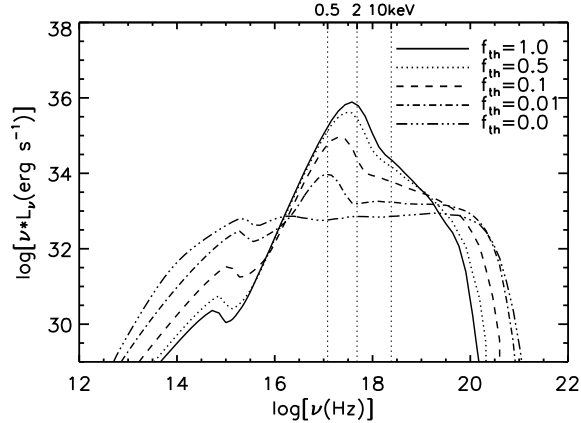}
\caption{\label{sp:fth} Emergent spectra of the ADAF around NSs 
for different $f_{\rm th}$ with $\dot m=5\times 10^{-3}$. } 
\end{figure*}

In the panel (1) of Fig. (\ref{fth}), we plot the ion temperature $T_{\rm i}$
and the electron temperature $T_{\rm i}$ of the ADAF as a function of radius 
for different $f_{\rm th}$ with $\dot m=5\times 10^{-3}$. As we can see, $T_{\rm i}$
nearly does not change with changing $f_{\rm th}$. While $T_{\rm e}$ increases 
with decreasing $f_{\rm th}$,  which can be easily understood as follows. A smaller value   
of $f_{\rm th}$ means that less energy of the internal energy stored in the ADAF and the 
radial kinetic energy of the  ADAF transfered onto the surface of the NS can be thermalized 
as the soft photons to cool the accretion flow, consequently predicting a relatively higher
$T_{\rm e}$. 
In the panel (2) of Fig. (\ref{fth}), we plot the 
Compton scattering optical depth $\tau_{\rm es}$ as a function of radius for NSs for different
$f_{\rm th}$ with $\dot m=5\times 10^{-3}$.
It is shown that there is a slight decrease of  $\tau_{\rm es}$ with decreasing $f_{\rm th}$.
In the panel (3) of  Fig. (\ref{fth}), we plot the Compton $y$-parameter as a function of 
radius for NSs for different $f_{\rm th}$ with $\dot m=5\times 10^{-3}$.
It is shown that the Compton $y$-parameter increases with decreasing $f_{\rm th}$, which predicts
a harder hard X-ray spectrum with decreasing $f_{\rm th}$, as can be seen in Fig. (\ref{sp:fth}).
In the  panel (4) of  Fig. (\ref{fth}),  we plot $\Omega/\Omega_{\rm K}$ 
as a function of radius for NSs for different $f_{\rm th}$ with $\dot m=5\times 10^{-3}$. 
It can be seen that $\Omega/\Omega_{\rm K}$ systematically decreases with decreasing $f_{\rm th}$, 
which can be understood as, with a decrease of $f_{\rm th}$, the radiative efficiency of 
the accretion flow decreases, resulting in a decrease of the angular velocity, as discussed in
Section \ref{sec_BH}. 
We list the radiative features of the ADAF around NSs for different $f_{\rm th}$ with 
$\dot m=5\times 10^{-3}$ in Table (\ref{fth_effect}).
We show that $L_{*}/L_{G}$ slightly changes with changing $f_{\rm th}$. 
It can be seen that the temperature of the thermal soft X-ray component decreases from $0.45$ keV 
to  $0.14$ keV for $f_{\rm th}$ decreasing from 1.0 to 0.01. Meanwhile, the  X-ray 
luminosity between 0.5 to 10 keV decreases from 
$1.0\times 10^{36}$ erg $\rm s^{-1}$ to $9.1\times 10^{33}$ erg $\rm s^{-1}$. 
For $f_{\rm th}=0.0$, there is no such a thermal soft X-ray component, and the emergent spectrum is
very similar to the case of BHs with the difference only from the mass between NSs and BHs. 

\begin{table*}
\caption{Radiative features of the ADAF around NSs 
for different $f_{\rm th}$.
$L_{*}/L_{\rm G}$ is the ratio of the energy of the ADAF transfered onto the 
surface of the NS per second to the accretion luminosity. 
$T_{*}$ is the effective temperature at the surface of NSs.
$L_{\rm 0.5-10 keV}$ is the luminosity between 0.5 and 10 $\rm keV$.}
\centering
\begin{tabular}{ccccccccc}
\hline\hline
$m$ &  $R_{*} \ ($\rm km$)$ & $\alpha$ & $\beta$ & $f_{\rm th}$ & $\dot m$  & $L_{*}/L_{\rm G}$  & $T_{*} \ (\rm keV)$ 
& $L_{\rm 0.5-10 keV}\ (\rm erg \ s^{-1}) $ \\
\hline
$1.4$ &12.5 & 0.3  &0.95   & 1.0    &   $5\times10^{-3}$   &56.0\%    &0.45    & $1.0\times 10^{36}$ \\
$1.4$ &12.5 & 0.3  &0.95   & 0.5    &   $5\times10^{-3}$   &57.3\%    &0.38    & $5.5\times 10^{35}$ \\
$1.4$ &12.5 & 0.3  &0.95   & 0.1    &   $5\times10^{-3}$   &59.1\%    &0.26    & $1.1\times 10^{35}$ \\
$1.4$ &12.5 & 0.3  &0.95   & 0.01   &   $5\times10^{-3}$   &60.3\%    &0.14    & $9.1\times 10^{33}$ \\
$1.4$ &12.5 & 0.3  &0.95   & 0.0    &   $5\times10^{-3}$   &60.5\%    &-       & $2.0\times 10^{33}$ \\
\hline\hline
\end{tabular}
\\
\label{fth_effect}
\end{table*}

\section{Discussions}
\subsection{On the interaction between the accretion flow and the NS}
In this paper, we just consider the radiative coupling between the ADAF and
the NS. Specifically, we consider that the internal energy stored in the ADAF and radial kinetic energy
of the ADAF are transfered onto the surface of the NS. Furthermore, we assume that only 
a fraction, $f_{\rm th}$, of this energy is thermalized at the surface of the NS as the soft 
photons to be scattered in the ADAF to self-consistently calculate the structure
of the ADAF. The interaction between the accretion flow and the NS is complicated, one of which is the 
boundary layer problem \citep[][for reivew]{Gilfanov2014}. 
As we have mentioned in the introduction part, there exists a critical mass accretion rate 
$\dot M_{\rm crit}$. For $\dot M \gtrsim \dot M_{\rm crit}$, the accretion 
flow will be in the form of a cool, geometrically thin, optically thick disc. In this case,
the key point for understanding the boundary layer problem is how the accreted matter 
in the disc is decelerated from its Keplerian orbital velocity about half the speed of light
to the NS's rotational velocity, in which $\sim$ half of the gravitational energy will be released
\citep{Popham1992,Narayan1993,Inogamov1999,Popham2001}.
However, we think that such a boundary layer problem is not serious for the accretion with 
$\dot M \lesssim \dot M_{\rm crit}$. As we know, generally, if $\dot M \lesssim \dot M_{\rm crit}$,
the accretion flow will transit from the cool, optically thick disc to the hot, optically thin  ADAF. 
Since ADAF is hot and optically thin, intrinsically the angular velocity of the ADAF is sub-Keplerian.
Because the angular velocity of the ADAF is relatively lower, as we can imagine, it 
is relatively easier to establish the equilibrium between ADAF and the NS. Meanwhile, because 
the angular velocity of the ADAF is relatively lower, the angular kinetic energy is not
important compared with the internal energy and the radial kinetic energy of the ADAF. 
In this paper, we simply ignore the contribution of the angular kinetic energy transfered onto the surface
of the NS as the soft luminosity to calculate the structure of the ADAF.

As has been shown in the panel (1), (2), (3) and (4) of Fig. \ref{fth},
a change of $f_{\rm th}$ can significantly change the structure of the ADAF. 
However, we do not know the detailed physics about the interaction between the energy transfered onto
the surface of the NS and the matter at the surface of the NS. 
Specifically, as an example, it is unclear how much of this energy can be thermalized as the soft 
photons for the Comptonization in the ADAF. Meanwhile, it is unclear how much of this energy is converted to 
other forms of energy, such as the rotational energy of the NS.  
As can be seen in Fig. \ref{sp:fth},    
a change of $f_{\rm th}$ can significantly change the emergent spectra of the ADAF. 
We also should keep in mind that, in the present paper we do not consider the effect of 
the remaining fraction, 1-$f_{\rm th}$, of the energy transfered onto the surface of the NS on
the emergent spectrum of the ADAF. As we mentioned, it is possible that such a fraction of the energy 
could be converted to the rotational energy of the NS. 

\citet[][]{Burke2018} found that there is a clear trend between the key Comptonization 
properties and the NS spin for a given accretion rate in the range of 
$L_{\rm X}/L_{\rm Edd} \backsim 0.005-0.1$. In the case of Newtonian approximation, the energy released 
in the boundary layer between the accretion flow and the NS for the Comptonization can be 
expressed as, 
\begin{eqnarray}\label{L_bl}
L_{\rm bl}=2\pi^2 \dot M R_{*}^2 (\nu_{*}-\nu_{\rm NS})^2,
\end{eqnarray} 
where $\nu_{\rm *}$ is the rotational frequency of the accretion flow at $R_{*}$, 
and $\nu_{\rm NS}$  is the rotational frequency of the NS. 
Here we would like to address that $L_{*}$ calculated from equation (\ref{soft_L}) always dominates
$L_{\rm bl}$ calculated from equation (\ref{L_bl}) in the framework of the  ADAF solution 
with $\dot m \lesssim 5\times 10^{3}$ for a wide 
range of $\nu_{\rm NS}$.  For example, $L_{*}$ is $8.1\times 10^{35} \ \rm erg \ s^{-1}$ 
for $\dot m=5\times 10^{-3}$. While 
$L_{\rm bl}$ is $7.7 \times 10^{34} \ \rm erg \ s^{-1}$ for $\dot m=5\times 10^{-3}$ 
with $\nu_{\rm NS}=0$, which is roughly one order of magnitude lower than that of $L_{*}$. 
Assuming $\nu_{\rm NS}=500$,  $L_{\rm bl}$ is $2.4\times 10^{31} \ \rm erg \ s^{-1}$, which 
is more than four orders of magnitude lower than that of  $L_{*}$.
So it can be seen that the effect of NS spin on the spectrum is always not important in the
ADAF case. However, we also need to note that, in the present paper, 
we indeed do not know how the NS spin can affect of the emergent spectrum of the ADAF if the 
full general relatively is considered, which will be studied in the not far future. 
Finally, we suggested that the value of $f_{\rm th}$ could be constrained by fitting 
the available X-ray data with high precisions, such as the X-ray data of 
$\it XMM$-$\it Newton$ in the range of 0.5-10 keV, which will be done in the future work for details.

\subsection{On the effect of large-scale magnetic field }
As we can see, in this paper, we do not consider the effect of the strong large-scale 
magnetic filed on the ADAF solution, which makes our solutions can only be applied to the  
NSs with a relatively weak magnetic filed.
Theoretically, if a strong large-scale magnetic field (typically $\gtrsim 10^{8-9} $ G) is 
existed around NSs, the geometry and the structure of the accretion flow can be 
changed. In this case, generally, there exists a critical radius $R_{\rm M}$, at which 
the magnetic pressure is equal to the ram and gas pressure of the matter in the accretion flow. 
At the region of $R>R_{\rm M}$, the matter of the accretion flow can roughly keep the symmetric 
structure in the angular direction. While at the region of $R<R_{\rm M}$, the matter in the accretion 
flow will be gradually controlled by the magnetic field, the symmetric structure will be disrupted, 
forming an accretion column \citep[][]{Frank2002}. The column accretion has very clear 
observational effects, i.e., the X-ray pulsar. We should note that, although it is generally believed 
that the strong magnetic filed can significantly alter the X-ray spectrum, the observational
evidence for the effect of the magnetic filed on the X-ray spectrum is still in debate.
\citet[][]{Wijnands2015} compared three accreting millisecond X-ray pulsars (AMXP), i.e., 
IGR J18245, NGC 6440 X-2, and IGR J00291+5934 with a sample composed of 11 low-level accreting
non-pulsating NSs, they did not find significant differences of the X-ray spectra 
in the range of 0.5-10 keV between the AMXP and the non-pulsating NSs. However, since the sample 
of AMXP in \citet[][]{Wijnands2015} is very small, the authors also addressed that they 
can not draw any strong conclusions, i.e., a dynamically important magnetic field can change 
the X-ray spectra of accreting NSXRBs significantly or not.

\subsection{On the effect of outflow}
In this paper, a constant mass accretion rate along the radial direction 
is assumed to calculate the structure of the ADAF around NSs.
In the history, the ADAF solution with a constant mass accretion rate was indeed applied to 
fit the broad-band SED of BH soft X-ray transient in quiescence \citep[e.g.][]{Narayan1996fit}. 
However, as the analysis in \citep[][]{Narayan1994}, ADAF solution has a positive Bernoulli 
parameter (the hot nature of the ADAF), which implies that the gas is not bound to the BH.
So it is suggested that ADAF solution is associated with strong outflows driven by the 
thermal pressure \citep[e.g.][]{Meier2001,Blandford1999}. Further, by fitting the broad-band 
SED of $\rm Sgr A^{*}$ from radio to X-rays with the  ADAF, 
it was found that the mass accretion rate decreases with decreasing radius with a formula of
$\dot m(r) \propto r^{0.3}$ \citep[][]{Yuan2003}, which was confirmed by the recent 
hydrodynamical and magnetohydrodynamical simulations around BHs \citep[][]{Yuan2012a,Yuan2012b}. 
Since the ADAF solution around NSs is also hot, as has been shown in this paper, 
we can imagine that the ADAF solution around NSs should be also associated with
outflows, resulting in a decrease of the mass accretion rate with decreasing radius.
If the mass accretion rate decreases with decreasing radius, both the 
structure and the emergent spectrum of the ADAF will change compared with a constant mass 
accretion rate, which of course will change the temperature of the thermal soft
X-ray component as we focus in this paper.
The dependence of the mass accretion rate $\dot m$ on the radius $r$ around NSs is unclear.
The radius-related mass accretion rate, such as the form of $\dot m(r) \propto r^{b}$,  
will be considered in the future work of the detailed spectral fitting for the value of $b$ 
around NSs.

\section{Conclusions}
In this paper, we investigate the origin of a thermal soft X-ray component detected 
in low-level accreting NSs based on the self-similar solution of the ADAF.  
We compare both the electron temperature $T_{\rm e}$ and the Compton $y$-parameter derived from the ADAF 
solution for BHs and NSs. It is found that both  $T_{\rm e}$ and $y$ of the ADAF around NSs are
systemically lower than that of BHs, which consequently predicts a softer hard X-ray spectrum 
for NSs compared with the case for BHs.
We find that the temperature of the thermal soft X-ray component decreases with decreasing mass 
accretion rate, which is qualitatively consistent with observations.
We test the effect of $f_{\rm th}$ on the structure and the emergent spectrum the ADAF,  
with a focus on the relationship between the temperature of the thermal soft X-ray component
and $f_{\rm th}$.  Specifically, the temperature of the thermal soft X-ray component decreases 
with decreasing $f_{\rm th}$. We address that in the current paper, we do not consider 
the effect of the strong large-scale magnetic filed on the ADAF, which makes that our solution 
can only be applied to NSs with a relatively weak magnetic filed. We discuss the importance of the
outflows on the emergent spectrum of the ADAF around NSs. 
Finally, We suggest that the detailed X-ray spectral fitting can help to constrain 
the model parameters, such as $f_{\rm th}$, in the future work. 


\section*{Acknowledgments}
We thank the anonymous referee for his/her very expert comments and suggestions. 
This work is supported by the National Natural Science Foundation of
China (Grants 11773037 and 11673026), the gravitational wave pilot B (Grants No. XDB23040100), 
and the National Program on Key Research and Development Project (Grant No. 2016YFA0400804).

\bibliographystyle{mnras}
\bibliography{qiaoel}


\bsp	
\label{lastpage}
\end{document}